\documentclass[conference,a4paper]{APSIPA2021}

\usepackage[dvipdfmx]{graphicx}
\usepackage{xcolor}
\usepackage[skip=2pt]{subcaption}
\usepackage{url}

\usepackage{array, amsmath, amssymb, amsfonts, bm, mleftright}
\usepackage{multirow, booktabs}
\usepackage{algorithm, algpseudocode}
\usepackage{xargs}
\usepackage{comment}

\usepackage{geometry}
\usepackage{setspace}

\begin{document}
\title{LEAD Dataset: How Can Labels for Sound Event Detection Vary Depending on Annotators?}
\author{
\authorblockN{
Naoki Koga\authorrefmark{1},
Yoshiaki Bando\authorrefmark{2} and
Keisuke Imoto\authorrefmark{1}
}

\authorblockA{
\authorrefmark{1}Doshisha University, Japan,
\authorrefmark{2}National Institute of Advanced Industrial Science and Technology, Japan \\
E-mail: keisuke.imoto@ieee.org}
}

\maketitle
\begin{abstract}
In this paper, we introduce a LargE-scale Annotator's labels for sound event Detection (LEAD) dataset, which is the dataset used to gain a better understanding of the variation in strong labels in sound event detection (SED).
In SED, it is very time-consuming to collect large-scale strong labels, and in most cases, multiple workers divide up the annotations to create a single dataset.
In general, strong labels created by multiple annotators have large variations in the type of sound events and temporal onset/offset.
Through the annotations of multiple workers, uniquely determining the strong label is quite difficult because the dataset contains sounds that can be mistaken for similar classes and sounds whose temporal onset/offset is difficult to distinguish.
If the strong labels of SED vary greatly depending on the annotator, the SED model trained on a dataset created by multiple annotators will be biased.
Moreover, if annotators differ between training and evaluation data, there is a risk that the model cannot be evaluated correctly.
To investigate the variation in strong labels, we release the LEAD dataset, which provides distinct strong labels for each clip annotated by 20 different annotators. The LEAD dataset allows us to investigate how strong labels vary from annotator to annotator and consider SED models that are robust to the variation of strong labels.
The LEAD dataset consists of strong labels assigned to sound clips from TUT Sound Events 2016/2017, TUT Acoustic Scenes 2016, and URBAN-SED. 
We also analyze variations in the strong labels in the LEAD dataset and provide insights into the variations. 
\end{abstract}

\section{Introduction}
\label{sec:intro}
Sound event detection (SED) \cite{Mesaros_TASLP2019_01,Mesaros_SPM2021_01} is a fundamental task in environmental sound analysis in which types, onsets, and offsets of sound events included in a sound clip are estimated. 
SED has various real-world applications, such as residential street surveillance \cite{Clavel_ICME2005_01} and semantic video search based on audio and visual content \cite{Bugalho_INTERSPEECH2009_01}. 
The standard approach to SED is to utilize a deep neural network (DNN) in a strongly supervised manner \cite{Cakir_IJCNN2015_01,Cakir_TASLP2017_01,Vaswani_NIPS2017_01}.

Examples of datasets used in strongly supervised SED are TUT Sound Events (SE) 2016, 2017 \cite{Mesaros_EUSIPCO2016_01,Mesaros_DCASE2017_01}, URBAN-SED \cite{Salamon_WASPAA2017_01,Salamon_ACMMM2014_01}, and AudioSet with temporally strong labels \cite{Gemmeke_ICASSP2017_01,Hershey_ICASSP2021_01}.
TUT SE 2016 and 2017 contain 47 sound clips in total with 3--5 minutes long and strong labels with 18 and 6 sound event classes, respectively.
TUT SE 2016 and 2017 show that annotating strong labels to sound events with relatively large signal lengths is very laborious.
URBAN-SED contains 10,000 sound clips with 10 seconds long and 50,000 annotated sound events.
To alleviate the labor involved in annotation, URBAN-SED has been constructed synthesizing independently recorded sound events.

\begin{figure}[t]
    \centering
    \hspace{0pt}
    \includegraphics[width=1.0\columnwidth]{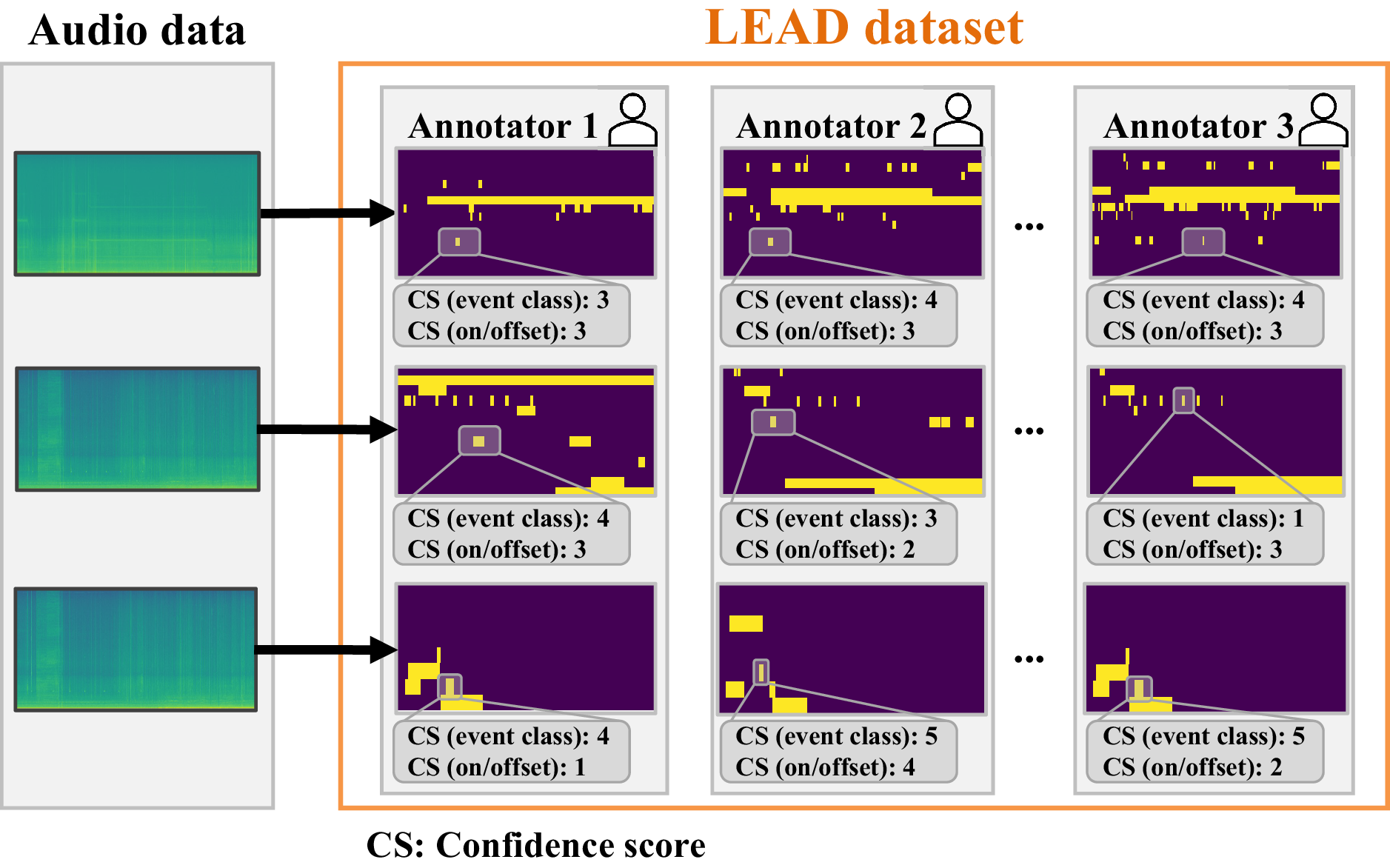}
    \vspace{-12pt}
    \caption{Overview of LEAD dataset}
    \label{fig:overview_LEAD}
    \vspace{-16pt}
\end{figure}

In SED, it is very laborious to collect large-scale strong labels, and in most cases, multiple workers divide up the annotations to create a single dataset.
Crowdsourcing is one of such methods used to assign a large number of strong labels to sound clips. 
However, previous works have shown that the strong labels created by multiple annotators tend to have a large variation in the type of sound events and temporal onset/offset \cite{MartinMorato_TASLP2023_01,Hershey_ICASSP2021_01}.
The strongly annotated dataset is used for not only the model training of strongly supervised methods but also the system evaluation of weakly supervised SED.
Thus, the variation in the annotation of strong labels is still a serious problem that should be addressed for SED.
Although many conventional studies have been conducted to investigate the variation in sound event labels, no previous studies have provided an open dataset for investigating the variation in strong labels.

To investigate the variation in strong labels and construct a robust method against the variation, we introduce the LEAD dataset as shown in Fig. \ref{fig:overview_LEAD}.
The LEAD dataset contains (i) strong labels that 20 annotators assigned to each sound clip in TUT SE 2016/2017, a part of TUT Acoustic Scenes (AS) 2016, and a part of URBAN-SED, (ii) a confidence score for selecting the sound event class to each sound event instance, and (iii) a confidence score for providing the onset/offset of each sound event instance.
We describe the steps and conditions for assigning strong labels and discuss in detail the variations in strong labels and confidence scores in the LEAD dataset. 
We then evaluate the impact of the temporal variation in the strong labels on the SED performance.
%
\section{Related Works}
\label{sec:related_work}
In some conventional works, the variation in sound event labels assigned to sound clips was investigated.
Fonseca et al. \cite{Fonseca_ICASSP2019_01} have released FSDnoisy18k, which is a dataset for sound event classification containing a small amount of clean data and large-scale real-world noisy data with weak sound event labels.
They reported that about 60\% of the data showed variation in sound event classes. 
Hershey et al. \cite{Hershey_ICASSP2021_01} have proposed a method in which sound event labels assigned by one annotator are modified by another annotator in sound event classification.
The experimental results showed that no agreement was reached between annotators after five attempts at this modifying process. 
Several studies on human auditory features that were not limited to specific tasks such as sound event classification have also been conducted.
Guastavino revealed that semantic features are more important than perceptual features when people perceive sounds \cite{Guastavino_Acustica2006_01}.

Some works have been conducted on generating more reliable labels using crowdsourced labels.
Cartwright et al. \cite{Cartwright_TOCHI2017_01} have investigated the trade-off between reliability and redundancy in the assignment of sound event labels by crowdsourcing. 
When the number of annotators exceeds five, the reliability of additional annotators begins to decrease.
Three findings were reported: first, when the number of annotators exceeds five, the value of additional annotators begins to decrease; second, the quality of labels improves by 90\% when the number of annotators is increased to 16; and third, the degree in variation of labels differs across classes.
M\'{e}ndez M\'{e}ndez et al. showed that when collecting sound event labels via a crowdsourcing platform, asking annotators to indicate their confidence level in each label assignment improves the quality of labels \cite{MendezMendez_PACMHCI2022_01}. 
Martin-Morat\'{o} et al. \cite{MartinMorato_TASLP2023_01} proposed a method to substitute the difficult task of annotating strong labels with easier tasks of annotating multiple weak labels to create more reliable strong labels in SED. 
Li and Liu \cite{Li_AAAI2015_01} proposed a method to select the optimal subset of annotators from a set of annotators when collecting labels by crowdsourcing, which utilizes a combinatorial optimization problem.
Bilen et al. \cite{Bilen_ICASSP2020_01} proposed an evaluation metric called the polyphonic sound detection score (PSDS) in SED, which is robust to the variation in strong labels among the annotators.
%
\section{LEAD Dataset}
\label{sec:LEAD_dataset}
We have constructed the LEAD dataset, which can be used to investigate the variation in strong event labels for SED among annotators.
The dataset consists of strong labels assigned to sounds in TUT Sound Events 2016/2017 \cite{Mesaros_EUSIPCO2016_01,Mesaros_DCASE2017_01}, a part of TUT Acoustic Scenes 2016 \cite{Mesaros_EUSIPCO2016_01}, and a part of URBAN-SED \cite{Salamon_WASPAA2017_01,Salamon_ACMMM2014_01}. 
For TUT Sound Events 2016/2017, we used all sound clips, which includes 2.9 h of sounds (47 files), whereas for TUT Acoustic Scenes 2016, we used a subset that includes all 15 acoustic scenes, which contains 2.6 h of sounds (314 files). 
For URBAN-SED, we used 0.17 h of sounds (60 files).
As URBAN-SED is the sound event dataset synthesized by Scaper \cite{Salamon_WASPAA2017_01}, the original dataset contains no variation among annotators. 
Moreover, in URBAN-SED, foreground sounds have a higher SNR than background sounds, and the annotators can more easily assign strong labels than in the case of using other datasets.
The LEAD dataset thus enables us to compare annotations for datasets with different characteristics. 
For each sound clip, 20 annotators assigned sound event labels and gave their onset/offset.
For TUT Sound Events 2016/2017 and TUT Acoustic Scenes 2016, we provided 20 candidate event labels for each acoustic scene, and the annotators selected the most suitable event labels from the candidates and then gave onsets/offsets. 
The annotators also assigned two confidence scores: one is the confidence score for selecting the type of sound event and the other is for providing the onset and offset of the sound event.
These confidence scores are attributed to each sound event instance. 
We set the confidence score on a five-point scale ranging from 1 (very unconfident) to 5 (very confident).
The annotators were asked to listen to an entire sound clip at least once during the annotation and to refer to a waveform plot and/or spectrogram if necessary.
The annotators were instructed to consider sound events occurring at intervals of more than 1 s as different sound events.

\begin{figure}[t!]
    \centering
    \begin{minipage}{0.49\hsize}
        \includegraphics[width=1.0\linewidth]{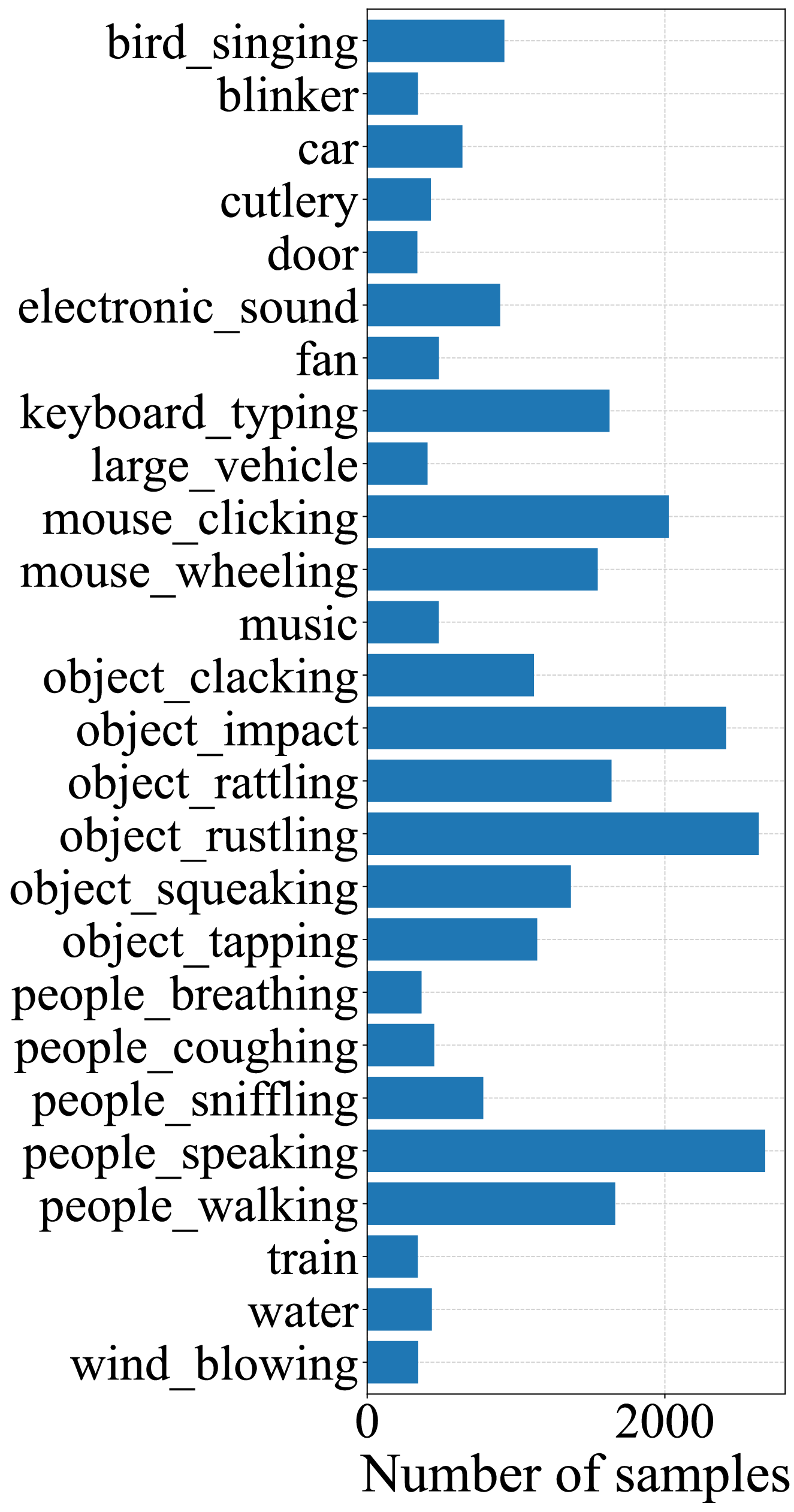}
        \subcaption{TUT Acoustic Scenes 2016}
        \label{fig:num_AS}
    \end{minipage}
    \begin{minipage}{0.49\hsize}
        \includegraphics[width=1.0\linewidth]{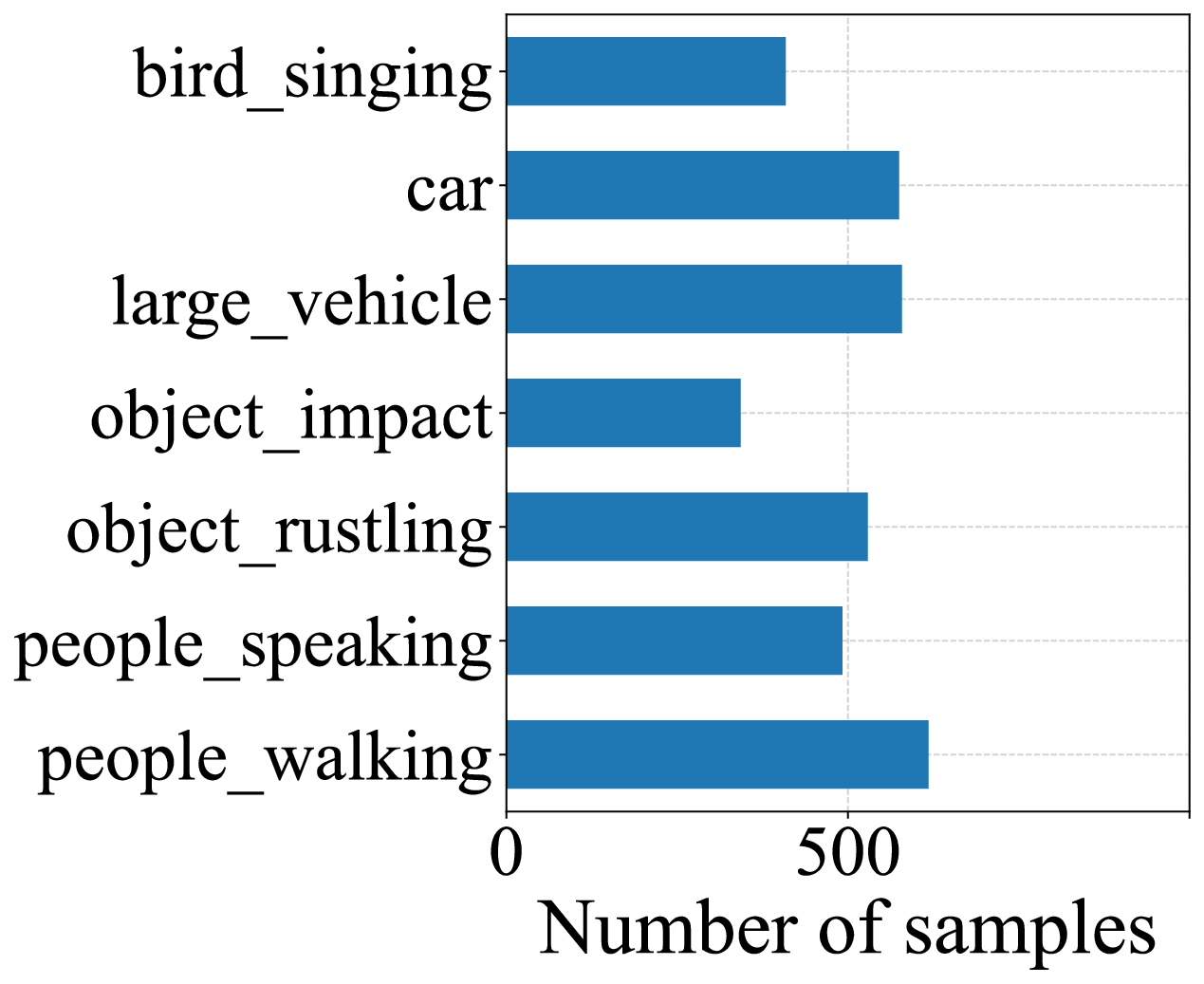}
        \subcaption{TUT Sound Events 2016/2017}
        \label{fig:num_SE}
        \vspace{6mm}
        \hspace*{3.2pt}
        \includegraphics[width=0.96\linewidth]{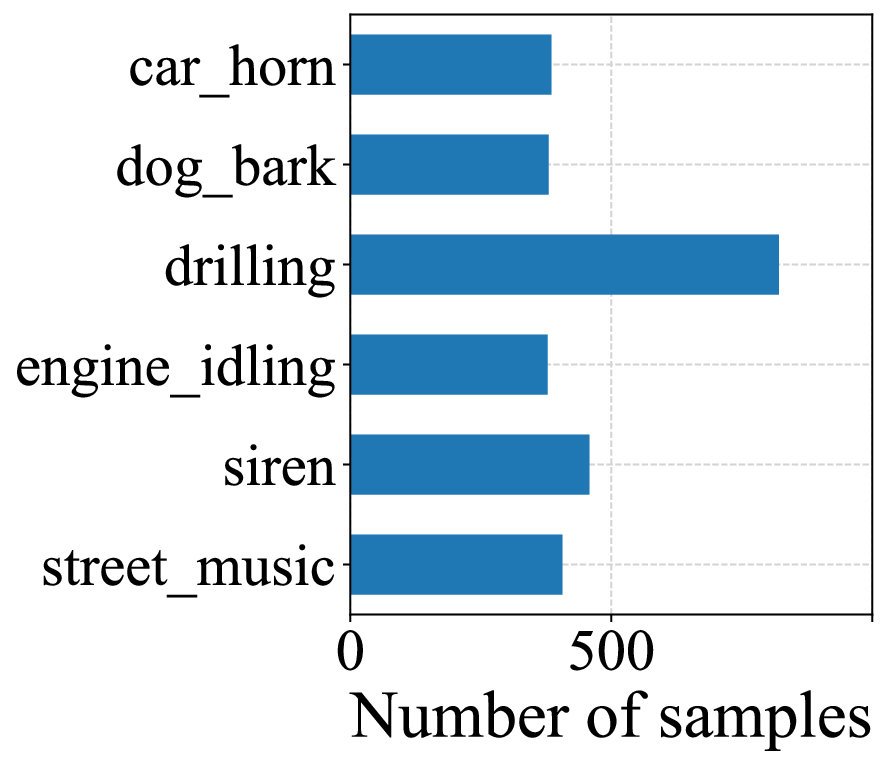}
        \subcaption{URBAN-SED}
        \label{fig:num_URBAN}
    \end{minipage}
    \vspace{3pt}
    \caption{Histograms of instances of sound events for excerpted classes in the LEAD dataset}
    \label{fig:num_instances}
\end{figure}

The annotations of the LEAD dataset are organized in the same tsv format as those of TUT Sound Events 2016/2017 and TUT Acoustic Scenes 2016, except for the confidence scores (CSs) to the sound event class and the onset/offset of each sound event instance, as follows.

\vspace{22pt}
\noindent \hspace{-5pt} \fbox{\vbox to46pt{\hsize247pt \vfil \ \vfil}}
\ \\

\vspace{-97pt}
\begin{tabular}{lllcc}
\!\!\!\!\!\!\multirow{2.05}{*}{\hspace{2.0pt} \textbf{Onset}} \!\!&\!\! \multirow{2.05}{*}{\hspace{3.0pt} \textbf{Offset}} & \multirow{2.05}{*}{\!\!\hspace{1.9pt} \textbf{Event class}} \!\!\!\!\!\!&\!\!\!\! \textbf{CS} \!\!&\!\!\!\!\! \textbf{CS}\\[-2pt]
\!\!\!\!\!\!& \!\!&\!\! \!\!\!\!\!&\!\!\!\! \textbf{(event)} \!\!&\!\!\!\!\! \textbf{(on/offset)}\\[5pt]
\!\!\!\!\!\!0.619375 \!\!&\!\! 12.304331 \!\!\!&\!\! car \!\!\!\!\!\!&\!\!\!\! 4 \!\!&\!\!\!\!\! 3\\
\!\!\!\!\!\!1.671731 \!\!&\!\! 224.798167 \!\!\!&\!\! bird singing \!\!\!\!\!\!&\!\!\!\! 5 \!\!&\!\!\!\!\! 4\\
\!\!\!\!\!\!3.926606 \!\!&\!\! 5.889908 \!\!\!&\!\! (object) impact \!\!\!\!\!\!&\!\!\!\! 3 \!\!&\!\!\!\!\! 3\\
\!\!\!\!\!\!7.616967 \!\!&\!\! 44.334651 \!\!\!&\!\! people walking \!\!\!\!\!\!&\!\!\!\! 4 \!\!&\!\!\!\!\! 4
\end{tabular}
\vspace{5pt}

\noindent The LEAD dataset is freely available online\footnote{https://github.com/KeisukeImoto/LEAD\_dataset}.

\begin{figure*}[t]
    \begin{minipage}{0.325\hsize}
        \centering
        \includegraphics[width=\hsize]{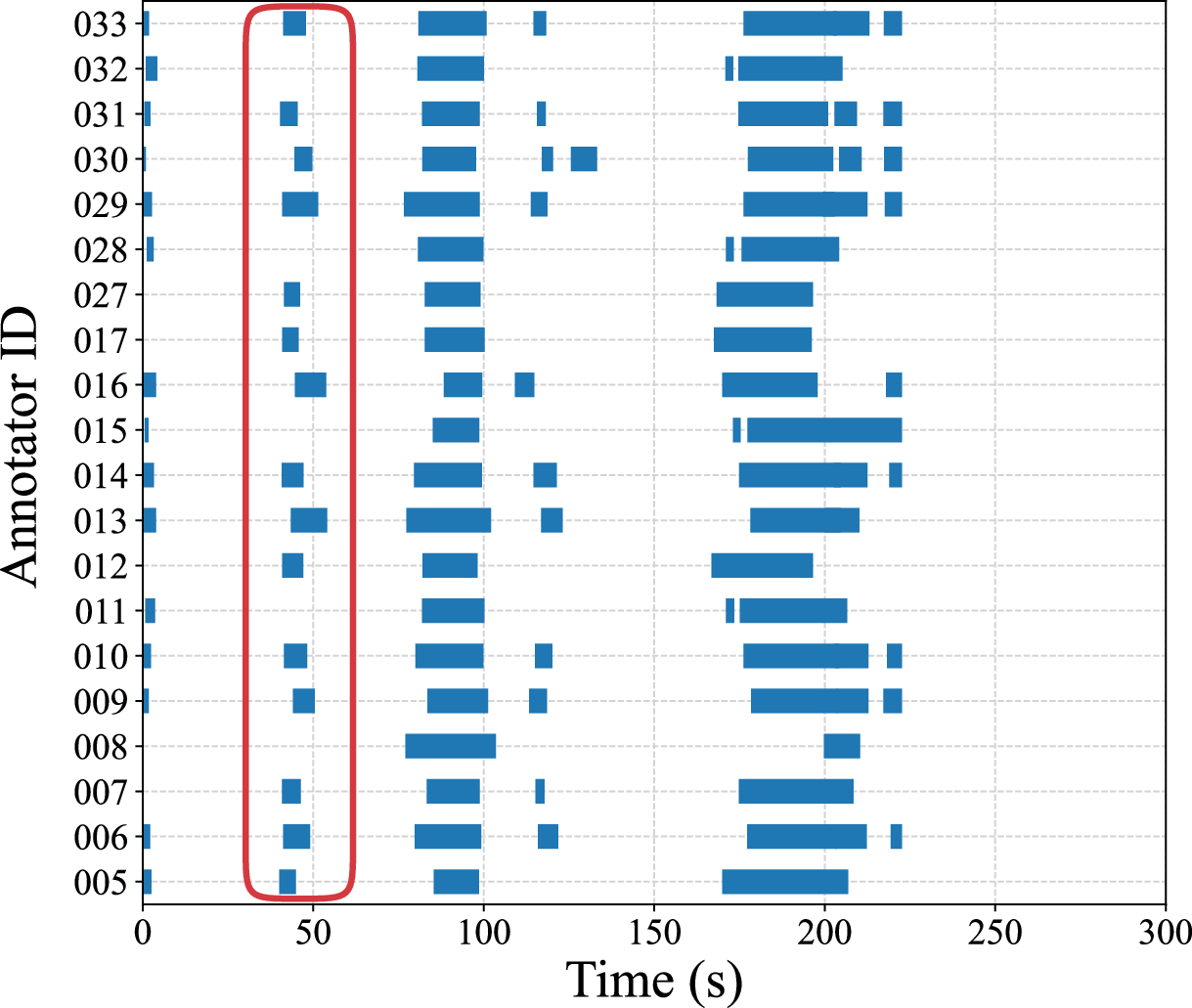}
        \caption{``bird\_singing'' in b006.wav}
        \label{fig:bird_singing}
    \end{minipage}
    \hfill
    \begin{minipage}{0.325\hsize}
        \centering
        \includegraphics[width=\hsize]{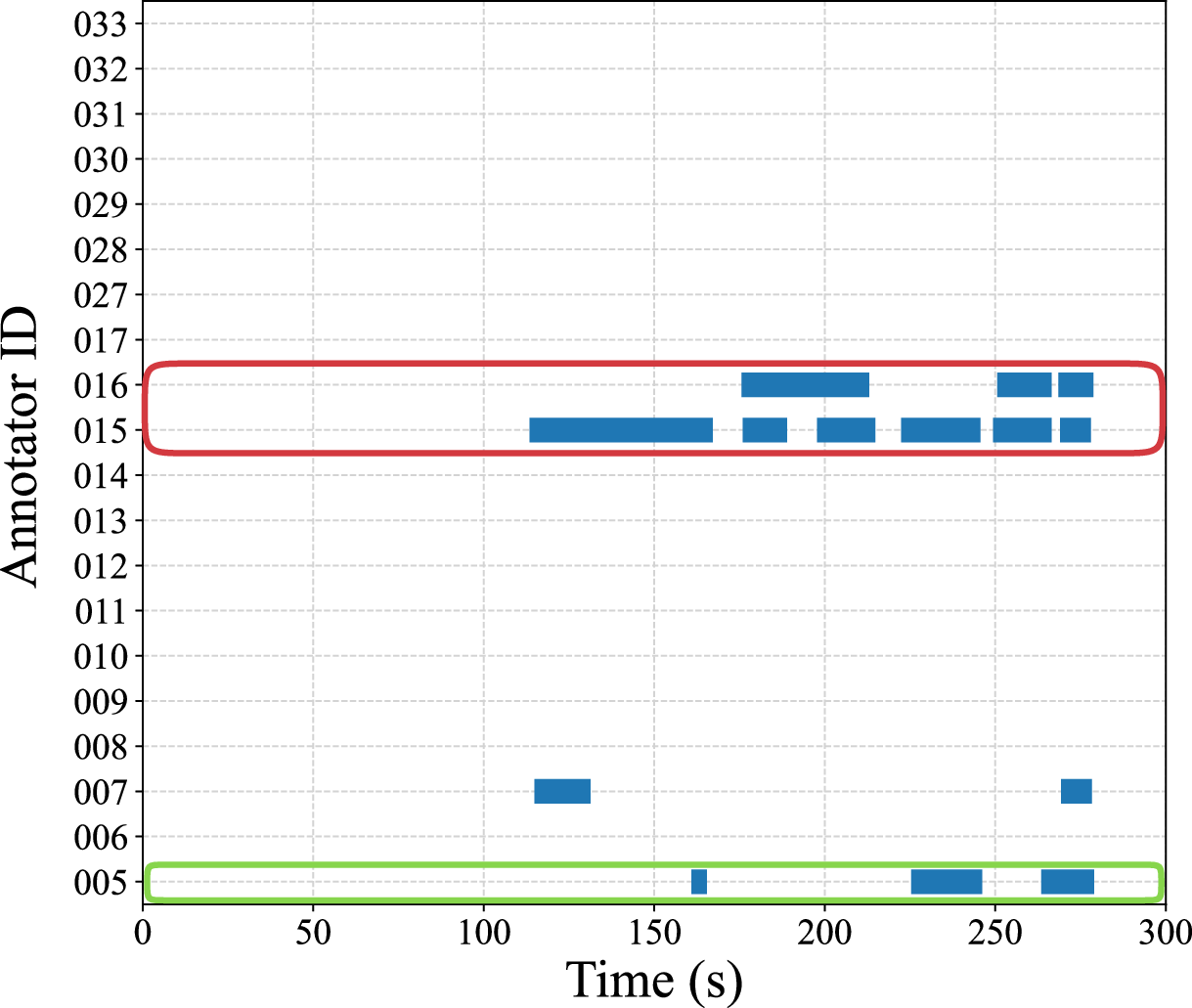}
        \caption{``water'' in b030.wav}
        \label{fig:b030_water}
    \end{minipage}
    \hfill
    \begin{minipage}{0.325\hsize}
        \centering
        \includegraphics[width=\hsize]{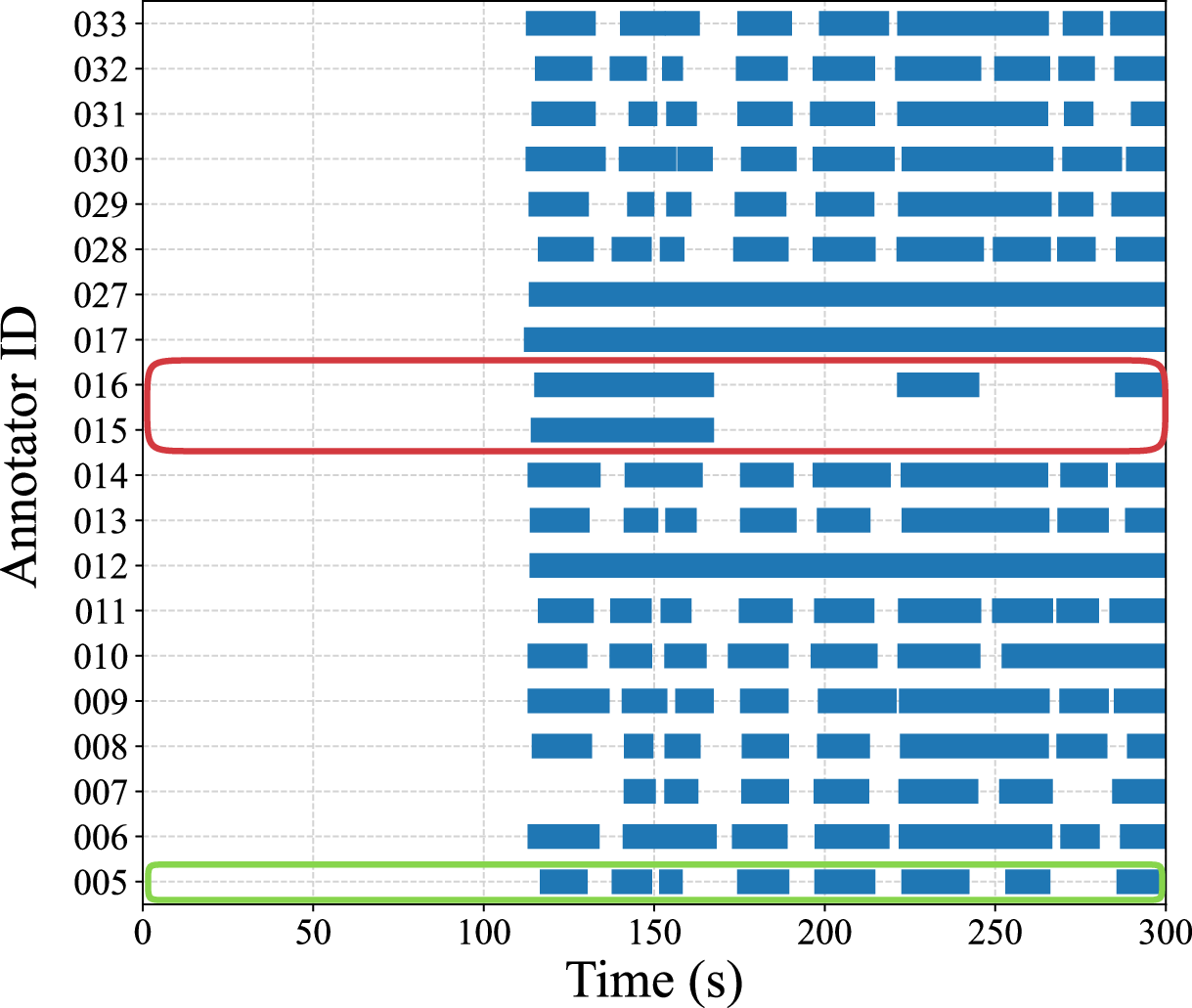}
        \caption{``washing\_dishes'' in b030.wav}
        \label{fig:washing_dishes}
    \end{minipage}
\end{figure*}

Figure \ref{fig:num_instances} shows the number of sound events for each event class collected through the annotation. 
Because of space limitations, only sound events with more than 300 instances are shown in Fig. \ref{fig:num_instances}. 
This indicates that the number of sound event instances and TUT AS 2016 is imbalanced, while URBAN-SED is well balanced since URBAN-SED is the synthesized dataset.

%
\section{Analysis of Variations In Strong Labels} 
\label{sec:var_labels}
In this section, we discuss possible variations in strong labels for SED by investigating the LEAD dataset.
Variations in sound event labels are of two categories: variation of sound event classes and time variation of onset/offset of sound event labels. 
Then, we investigate confidence scores of annotation for the sound event category and onset/offset of sound event labels.

\subsection{Variation in Sound Event Classes}
\label{ssec:event_classes_var}
\noindent {\bf Label deletion}\ \ \ 
Label deletion indicates that an event label is not assigned to an expected sound event.
If label deletion is present in the training data, a trained model might overlook sound events that should be detected.
Moreover, if label deletion occurs in the evaluation data, the evaluation results may include improper false positive samples.
Figure \ref{fig:bird_singing} shows an example of label deletion of the ``bird\_singing'' event in the LEAD dataset.
At around 50 seconds, we can see that annotators 008, 011, 015, 028, and 032 do not assign to an expected sound event.
One countermeasure against label deletion is to use a majority vote for strong labels among annotators. 
Another reliable solution is to estimate the annotator's competence using the Multi-Annotator Competence Estimation (MACE) \cite{Hovy_NAACLHLT2013_01} and then take a weighted average of the annotators' opinions. \\
%
\noindent {\bf Label insertion}\ \ \ 
Label insertion means assigning an event label to a sound event that does not exist.
If label insertion occurs in the training data, the trained model may detect false positive samples. 
On the other hand, improper false negative samples are included in the evaluation results if label insertion occurs in the evaluating data.
Figure \ref{fig:b030_water} shows example of label insertion in the LEAD dataset, in which annotator 005 gives ``water'' and ``washing\_dishes'' labels that probably do not exist.
One way to address label insertion is to use a majority vote among the annotators. \\
%
\noindent {\bf Label substitution}\ \ \ 
Label substitution refers to a situation in which a sound event class is confused with another similar category. 
One factor that causes label insertion is the existence of sound events that are auditorily similar but have different meanings.
For example, the sound of a ventilation fan and the sound of a car passing by are auditorily similar sound events but have different meanings.
Label substitution in the training dataset results in a mistrained model, which is likely to output cross-triggered results. 
Moreover, label substitution in the evaluation dataset causes a cross-trigger during system evaluation. 
A cross-trigger during system evaluation is better than a cross-trigger during model training but still disrupts system development in SED. 
Figures \ref{fig:b030_water} and \ref{fig:washing_dishes} show examples of label substitution, in which annotators 015 and 016 have confused sound events ``water'' and ``washing\_dishes''.  
To avoid label substitution when constructing a dataset, it is important not to include label candidates that are easily confused in the label list.
\\
%
\noindent {\bf Hierarchical label substitution}\ \ \ 
Hierarchical label substitution represents label substitution for a sound event ontology \cite{Nakatani_AAAI1998_01,Gemmeke_ICASSP2017_01}. 
Since the LEAD dataset (including TUT Acoustic Scenes 2016/2017, TUT Sound Events 2016/2017, and URBAN-SED) do not use ontology-based label candidates, hierarchical label substitution does not occur explicitly. 
However, the label candidates may implicitly have the hierarchical label structure; thus, it is important to carefully select the candidates and determine the procedure for selecting which hierarchical level of sound events should be selected.

\begin{table*}[t]
    \centering
    \small
    \caption{Average confidence scores on sound event class and onset/offset in LEAD dataset}
    \label{table:conf_score}
    \newcommand{\apsipaconfig}[1]{%
                \setlength\tabcolsep{1.2mm}
                   #1
        }
    \apsipaconfig{ 
        \begin{tabular}{lccccc}
            \toprule
            Dataset & TUT SE 2016 & TUT SE 2017 & TUT AS 2016 & URBAN-SED \\
            \midrule
            Confidence score on sound event class & 3.94 $\pm$ 0.16 & 4.00 $\pm$ 0.17  & 3.94 $\pm$ 0.12 & 4.03 $\pm$ 0.21 \\ 
            Confidence score on onset/offset & 4.04 $\pm$ 0.12 & 4.00 $\pm$ 0.19 & 4.11 $\pm$ 0.11 & 4.13 $\pm$ 0.13 \\
            \bottomrule
        \end{tabular}
    }
\end{table*}

\noindent {\bf Label integration}\ \ \ 
Label integration indicates that when multiple sound events with short duration occur in succession, only one integrated sound event label is annotated. 
Label integration is not necessarily a serious error.
However, if it is inconsistent between the training and evaluation data, the evaluation result contains many inappropriate false positive or false negative samples, making it difficult to correctly evaluate the performance of the trained model.
As an example of label integration, in Fig. \ref{fig:washing_dishes}, the sound event labels of annotators 012, 017, and 027 were integrated. 
As explained in Sec. \ref{sec:LEAD_dataset}, the annotators were instructed to treat sound events that occurred at intervals longer than 1 second separately.
Nevertheless, a few annotations include label integration, and it is important to have a common agreement in advance on how independently successive sound events are annotated.
\subsection{Temporal Variation of Strong Labels}
\label{ssec:temporal_var}
The temporal onset and offset of sound events also vary greatly depending on the annotator \cite{MartinMorato_TASLP2023_01,Hershey_ICASSP2021_01}. 
For example, in Fig. \ref{fig:bird_singing}, the unbiased standard deviation of the onset and offset are 2.87 and 1.39 s, respectively.
Here, sound events where label integration has occurred are excluded from the calculation of standard deviation.
The temporal variation in strong labels is inevitable, but sufficient solutions have not been found in previous studies \cite{MartinMorato_TASLP2023_01,Hershey_ICASSP2021_01}.
This not only affects the model training in SED, but also has a significant impact on the performance evaluation of the trained model.
For example, in DCASE 2020--2024 Challenge task 4, a 0.2 s collar is used in event-based evaluation metrics \cite{Turpault_DCASE2019_01,Ronchini_DCASE2021_01}.
However, in the annotations shown in Fig. \ref{fig:bird_singing}, the time variation in the onset and offset far exceeded 0.2 s.
This indicates that time variation can affect the evaluation results and the collar setting can be improved.
To mitigate the negative impact of time variation, one approach is to use an evaluation metric that is robust to the time fluctuation in strong labels, such as PSDS \cite{Bilen_ICASSP2020_01}.
%
\subsection{Confidence score for sound event label and onset/offset of sound event}
\label{ssec:feature_LEAD}

(ii) a confidence score for selecting the sound event class to each sound event instance, and (iii) a confidence score for providing the onset/offset of each sound event instance.

Table \ref{table:conf_score} shows the average confidence scores for selecting the sound event class to each sound event instance and for providing the onset/offset of each sound event instance with 95\% confidence intervals.
As URBAN-SED is the synthesized dataset, we expected that its confidence scores for both event labels and onset/offset would be high, however there is no significant difference from other datasets.
We also found that the average length of sound events may not significantly affect the confidence scores. 
For example, for ``people\_breathing'' (1.24 s), ``cupboard'' (3.99 s), and ``people\_shouting'' (12.04 s), the confidence scores for the type and onset/offset of sound events were unaffected by the sound event length.
%
\section{Experiments}
\label{sec:exp}
\subsection{Experimental Conditions}
\label{ssec:exp_condition}
We conducted experiments using the LEAD dataset to evaluate how the variation in strong labels affects the SED performance.
In this experiment, we evaluated a pseudo-SED system in which one of the 20 annotations was considered an output of the optimal SED system, and the remaining 19 annotations were considered the ground truth labels.
We calculated $20 \times 19 = 380$ pseudo-detection results in the same way and evaluated the SED performance. 
As evaluation metrics, we calculated the segment-based, event-based \cite{Mesaros_AppSci2016_01}, and intersection-based \cite{Bilen_ICASSP2020_01} micro-F-score.
The detailed parameter settings are shown in Table~\ref{table:conditions}. 

\begin{table}[tbp]
\centering
\small
\caption{Experimental conditions} 
\vspace{-4pt}
\label{table:conditions}
\begin{tabular}{lcc}
\toprule
Segment length (segment-based F-score) & \multicolumn{2}{c}{1.0 s} \\[1pt]
Collar (event-based F-score)   & \multicolumn{2}{c}{0.20 s}\\[-1pt]
$\rho_{\rm GTC}$, $\rho_{\rm DTC}$ & \multicolumn{2}{c}{\multirow{2}{*}{0.1, 0.1}} \\[-0.5pt]
(intersection-based F-score) \cite{Bilen_ICASSP2020_01}& \\
\bottomrule
\\
\end{tabular}
\small
\vspace{8pt}
\caption{SED performance in terms of micro F-score} 
\vspace{-4pt}
\label{table:performance}
\footnotesize
\begin{tabular}{cccc}
\toprule
\multirow{2}{*}{Dataset} & Segment-based & Event-based & Intersection-based \\
& micro-F-score & micro-F-score & micro-F-score \\
\midrule
TUT SE 2016 & 68.21\% & 8.17\% & 53.87\% \\
TUT SE 2017 & 61.79\% & 5.33\% & 32.59\% \\
TUT AS 2016 & 68.57\% & 32.59\% & 54.25\% \\
URBAN-SED & 72.76\% & 50.51\% & 40.52\% \\
\bottomrule
\end{tabular}
\vspace{14pt}
\caption{Example of SED performance} 
\vspace{-4pt}
\label{table:performance02}
\begin{tabular}{cccc}
\toprule
\multirow{2}{*}{Data name} & Segment-based & Event-based & Intersection-based \\
& micro-F-score & micro-F-score & micro-F-score \\
\midrule
b006.wav & 74.09\% &7.03\% & 75.73\% \\
a031.wav & 56.69\% & 23.06\% & 65.58\% \\
\bottomrule
\end{tabular}
\end{table}

\subsection{SED Performance}
\label{ssec:exp_result}
Table~\ref{table:performance} shows the SED performance for TUT SE 2016/2017, TUT AS 2016, and URBAN-SED in terms of segment-based, event-based, and intersection-based micro-F-scores. 
Considering that the experimental result does not include the detection errors of the SED model, the fact that most SED performances are below 70\% is an undesirable result.
In particular, the event-based F-score is quite low.
This result also supports that the idea the collar setting should be reconsidered.

Table~\ref{table:performance02} shows an example of the SED performance for b006.wav and a031.wav in TUT Sound Events 2016. 
In b006.wav, the sound of ``bird\_singing'' is dominant, and the temporal variation in onset/offset is very large, leading to a decrease in the SED performance in terms of the event-based metric. 
\subsection{Investigation of SED Performance Dependence on Collar Setting}
\label{ssec:add_exp_result}
\begin{figure}[t]
    \centering
    \includegraphics[width=0.90\hsize]{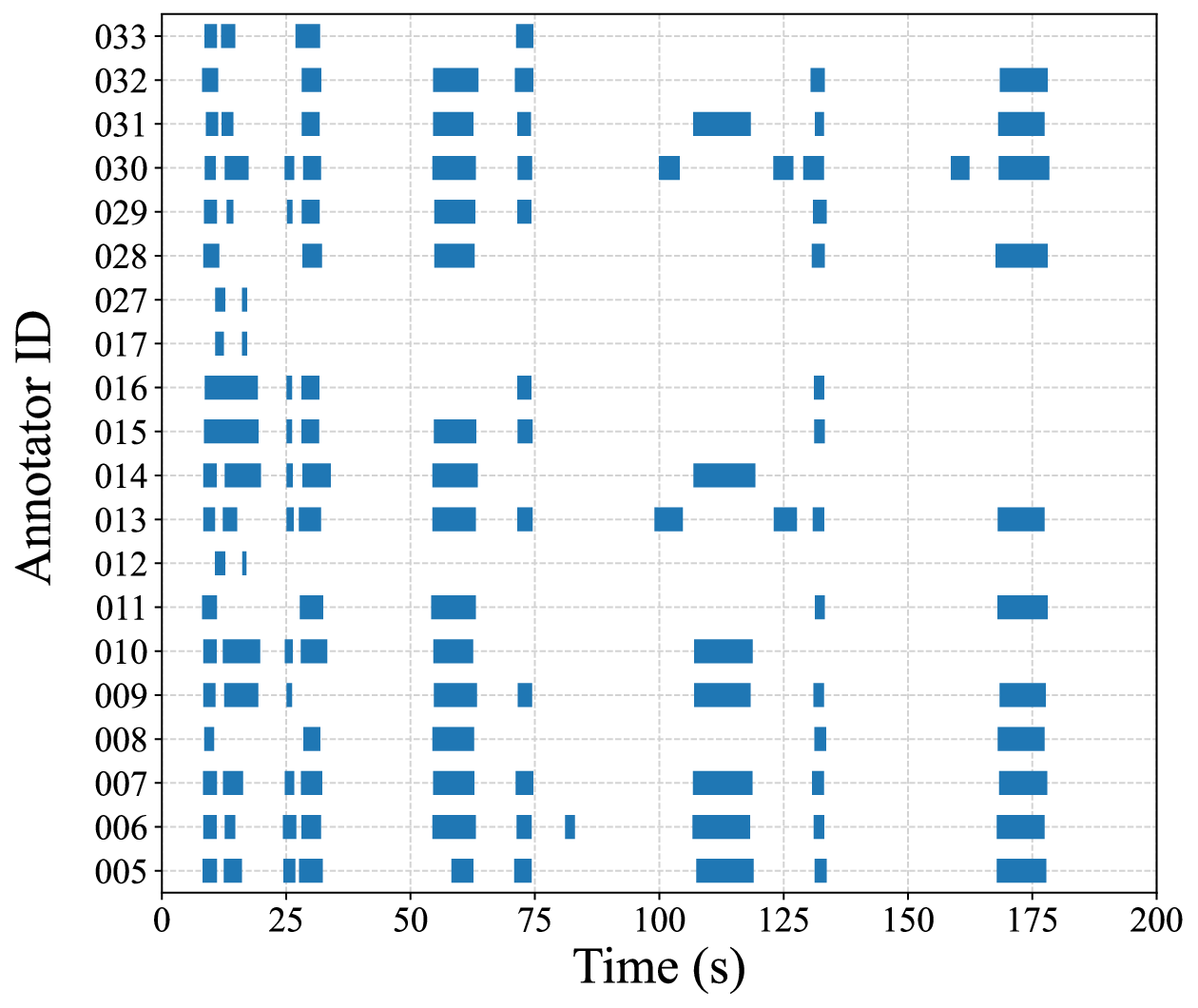}
    \vspace{-5pt}
    \caption{``Water'' labels in a031.wav}
    \label{fig:a031_water}
    \vspace{10pt}
    \centering
    \includegraphics[width=0.90\hsize]{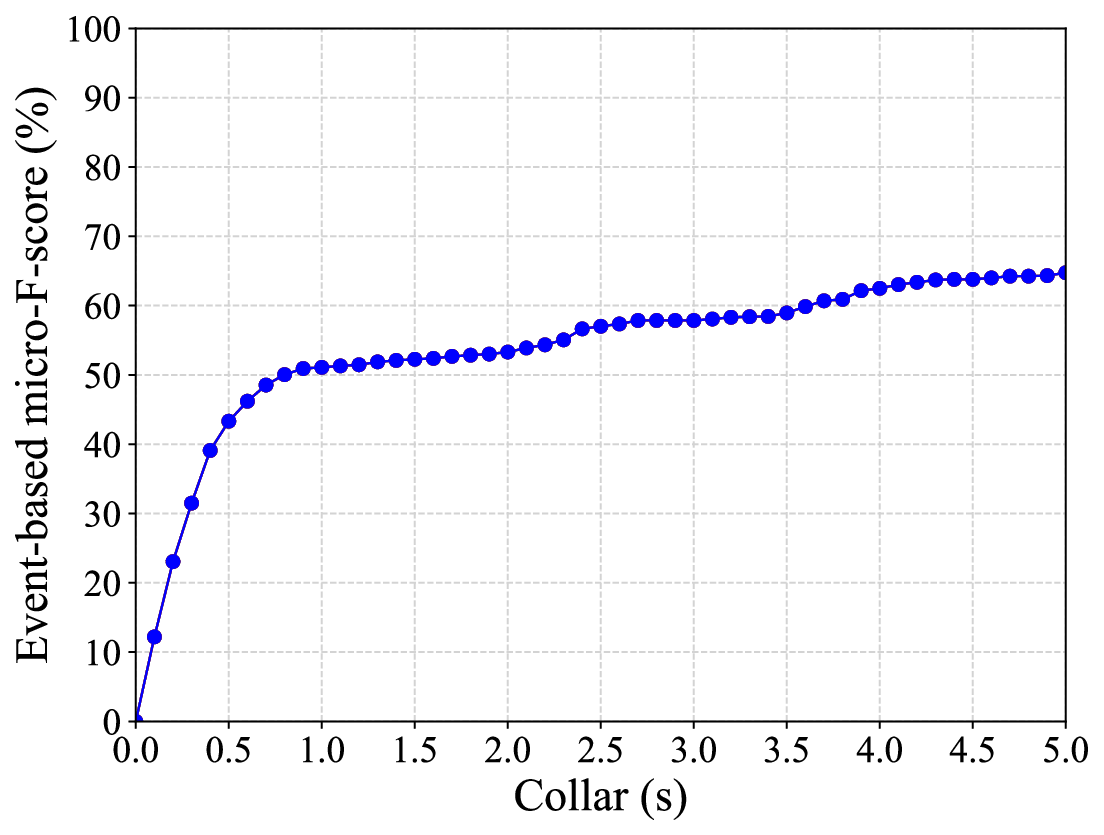}
    \vspace{-5pt}
    \caption{Event-based micro-F-score performance for various collar ratios}
    \label{fig:eventF1_collar}
\end{figure}

We then investigated the effects of temporal variation in sound event labels and the collar setting on SED performance. 
In this experiment, we evaluated the SED performance with various collar settings of the event-based metric using ``water'' sound in a031.wav.
Figure \ref{fig:a031_water} shows the annotation result of ``water'' in a031.wav. 
The collars were set at intervals of 0.1 s from 0.0 to 5.0 s.
As shown in Fig. \ref{fig:eventF1_collar}, the experimental results show that the event-based micro-F-score rapidly increases up to around 1.0 s. 
On the other hand, the unbiased standard deviations of onset and offset in water sounds in a031.wav are 3.14 and 3.73 s, respectively.
These results indicate that if the collar is set too short, the error in the performance evaluation result caused by the time variation in the sound event labels will be large.
For example, in DCASE 2022 Challenge task 4 \cite{Ronchini_DCASE2021_01}, the collar is set to 0.2 s as in the experiment in Sec. \ref{ssec:exp_result}. 
These results may be affected by the temporal variation in sound event labels, and the collar setting in SED performance evaluation should be reconsidered. 
\section{Conclusion}
\label{sec:concl}
In this paper, we introduced the LEAD dataset, where 20 annotators assigned strong labels to each sound clip.
We then investigated the variation in strong labels using the LEAD dataset, where the variation in strong labels is organized into the variation in sound event classes and the temporal variation in strong labels.
We also conducted experiments using the LEAD dataset to evaluate the effect of the variation in strong labels on the SED performance.
The experimental results show that the temporal variation in sound event labels seriously affects the SED performance: this could lower the SED performance by 30--40\% points.
Moreover, the experimental results showed that the short collar on the event-based F-score may affect the SED performance.
We believe that the LEAD dataset will accelerate research into the variation in strong labels in SED.

\section*{Acknowledgment}
This work was supported by JSPS KAKENHI Grant Numbers 22H03639 and 23K16908.


\bibliographystyle{IEEEtran}
\bibliography{mybib}

\begin{thebibliography}{10}
\providecommand{\url}[1]{#1}
\csname url@samestyle\endcsname
\providecommand{\newblock}{\relax}
\providecommand{\bibinfo}[2]{#2}
\providecommand{\BIBentrySTDinterwordspacing}{\spaceskip=0pt\relax}
\providecommand{\BIBentryALTinterwordstretchfactor}{4}
\providecommand{\BIBentryALTinterwordspacing}{\spaceskip=\fontdimen2\font plus
\BIBentryALTinterwordstretchfactor\fontdimen3\font minus
  \fontdimen4\font\relax}
\providecommand{\BIBforeignlanguage}[2]{{%
\expandafter\ifx\csname l@#1\endcsname\relax
\typeout{** WARNING: IEEEtran.bst: No hyphenation pattern has been}%
\typeout{** loaded for the language `#1'. Using the pattern for}%
\typeout{** the default language instead.}%
\else
\language=\csname l@#1\endcsname
\fi
#2}}
\providecommand{\BIBdecl}{\relax}
\BIBdecl

\bibitem{Mesaros_TASLP2019_01}
A.~Mesaros, A.~Diment, B.~Elizalde, T.~Heittola, E.~Vincent, B.~Raj, and
  T.~Virtanen, ``Sound event detection in the {DCASE} 2017 challenge,''
  \emph{IEEE/ACM Transactions on Audio, Speech, and Language Processing},
  vol.~27, no.~6, pp. 992--1006, 2019.

\bibitem{Mesaros_SPM2021_01}
A.~Mesaros, T.~Heittola, T.~Virtanen, and M.~D. Plumbley, ``Sound event
  detection: A tutorial,'' \emph{IEEE Signal Processing Magazine}, vol.~38,
  no.~5, pp. 67--83, 2021.

\bibitem{Clavel_ICME2005_01}
C.~Clavel, T.~Ehrette, and G.~Richard, ``Events detection for an audio-based
  surveillance system,'' \emph{Proc. IEEE International Conference on
  Multimedia and Expo (ICME)}, pp. 1306--1309, 2005.

\bibitem{Bugalho_INTERSPEECH2009_01}
M.~Bugalho, J.~Portelo, I.~Trancoso, T.~Pellegrini, and A.~Abad, ``Detecting
  audio events for semantic video search,'' \emph{Proc. INTERSPEECH}, pp.
  1151--1154, 2009.

\bibitem{Cakir_IJCNN2015_01}
E.~\c{C}akir, T.~Heittola, H.~Huttunen, and T.~Virtanen, ``Polyphonic sound
  event detection using multi label deep neural networks,'' \emph{Proc.
  International Joint Conference on Neural Networks (IJCNN)}, pp. 1--7, 2015.

\bibitem{Cakir_TASLP2017_01}
E.~\c{C}akir, G.~Parascandolo, T.~Heittola, H.~Huttunen, and T.~Virtanen,
  ``Convolutional recurrent neural networks for polyphonic sound event
  detection,'' \emph{IEEE/ACM Transactions on Audio, Speech, and Language
  Processing}, vol.~25, no.~6, pp. 1291--1303, 2017.

\bibitem{Vaswani_NIPS2017_01}
A.~Vaswani, N.~Shazeer, N.~Parmar, J.~Uszkoreit, L.~Jones, A.~Gomez, L.~Kaiser,
  and I.~Polosukhin, ``Attention is all you need,'' \emph{Advances in Neural
  Information Processing Systems}, vol.~30, pp. 1--11, 2017.

\bibitem{Mesaros_EUSIPCO2016_01}
A.~Mesaros, T.~Heittola, and T.~Virtanen, ``{TUT} database for acoustic scene
  classification and sound event detection,'' \emph{Proc. European Signal
  Processing Conference (EUSIPCO)}, pp. 1128--1132, 2016.

\bibitem{Mesaros_DCASE2017_01}
A.~Mesaros, T.~Heittola, A.~Diment, B.~Elizalde, A.~Shah, E.~Vincent, B.~Raj,
  and T.~Virtanen, ``Challenge setup: Tasks, datasets and baseline system,''
  \emph{Proc. Workshop on Detection and Classification of Acoustic Scenes and
  Events (DCASE)}, pp. 85--92, 2017.

\bibitem{Salamon_WASPAA2017_01}
J.~Salamon, D.~MacConnell, M.~Cartwright, P.~Li, and J.~P. Bello, ``Scaper: A
  library for soundscape synthesis and augmentation,'' \emph{Proc. IEEE
  Workshop on Applications of Signal Processing to Audio and Acoustics
  (WASPAA)}, pp. 344--348, 2017.

\bibitem{Salamon_ACMMM2014_01}
J.~Salamon, C.~Jacoby, and J.~P. Bello, ``A dataset and taxonomy for urban
  sound research,'' \emph{Proc. ACM International Conference on Multimedia
  (ACMMM)}, no.~4, pp. 1041--1044, 2014.

\bibitem{Gemmeke_ICASSP2017_01}
J.~F. Gemmeke, D.~P.~W. Ellis, D.~Freedman, A.~Jansen, W.~Lawrence,
  R.~Channing~Moore, M.~Plakal, and M.~Ritter, ``Audio set: An ontology and
  human-labeled dataset for audio events,'' \emph{Proc. IEEE International
  Conference on Acoustics, Speech and Signal Processing (ICASSP)}, pp.
  776--780, 2017.

\bibitem{Hershey_ICASSP2021_01}
S.~Hershey, D.~P.~W. Ellis, E.~Fonseca, A.~Jansen, C.~Liu, R.~Channing~Moore,
  and M.~Plakal, ``The benefit of temporally-strong labels in audio event
  classification,'' \emph{Proc. IEEE International Conference on Acoustics,
  Speech and Signal Processing (ICASSP)}, pp. 366--370, 2021.

\bibitem{MartinMorato_TASLP2023_01}
I.~Mart\'{i}n-Morat\'{o} and A.~Mesaros, ``Strong labeling of sound events
  using crowdsourced weak labels and annotator competence estimation,''
  \emph{IEEE/ACM Transactions on Audio, Speech, and Language Processing},
  vol.~31, pp. 902--914, 2023.

\bibitem{Fonseca_ICASSP2019_01}
E.~Fonseca, M.~Plakal, D.~P.~W. Ellis, F.~Font, X.~Favory, and X.~Serra,
  ``Learning sound event classifiers from web audio with noisy labels,''
  \emph{Proc. IEEE International Conference on Acoustics, Speech and Signal
  Processing (ICASSP)}, pp. 21--25, 2019.

\bibitem{Guastavino_Acustica2006_01}
C.~Guastavino, ``The ideal urban soundscape: Investigating the sound quality of
  french cities,'' \emph{Acta Acustica united with Acustica}, vol.~92, pp.
  945--951, 2006.

\bibitem{Cartwright_TOCHI2017_01}
M.~Cartwright, A.~Seals, J.~Salamon, A.~Williams, S.~Mikloska, D.~MacConnell,
  E.~Law, J.~P. Bello, and O.~Nov, ``Seeing sound: Investigating the effects of
  visualizations and complexity on crowdsourced audio annotations,'' \emph{ACM
  Transactions on Computer-Human Interaction}, vol.~1, no.~29, pp. 1--21, 2017.

\bibitem{MendezMendez_PACMHCI2022_01}
A.~E. M\'{e}ndez~M\'{e}ndez, M.~Cartwright, J.~P. Bello, and O.~Nov,
  ``Eliciting confidence for improving crowdsourced audio annotations,''
  \emph{Proc. ACM Human-Computer Interaction}, vol.~6, no.~88, pp. 1--25, 2022.

\bibitem{Li_AAAI2015_01}
H.~Li and Q.~Liu, ``Cheaper and better: Selecting good workers for
  crowdsourcing,'' \emph{Proc. the Association for the Advancement of
  Artificial Intelligence (AAAI)}, pp. 20--21, 2015.

\bibitem{Bilen_ICASSP2020_01}
C.~Bilen, G.~Ferroni, F.~Tuveri, J.~Azcarreta, and S.~Krstulovi\'{o}, ``A
  framework for the robust evaluation of sound event detection,'' \emph{Proc.
  IEEE International Conference on Acoustics, Speech and Signal Processing
  (ICASSP)}, pp. 61--65, 2020.

\bibitem{Hovy_NAACLHLT2013_01}
D.~Hovy, T.~Berg-Kirkpatrick, A.~Vaswani, and E.~Hovy, ``Learning whom to trust
  with {MACE},'' \emph{Proc. the Conference of the North {A}merican Chapter of
  the Association for Computational Linguistics: Human Language Technologies},
  pp. 1120--1130, 2013.

\bibitem{Nakatani_AAAI1998_01}
T.~Nakatani and H.~G. Okuno, ``Sound ontology for computational auditory scence
  analysis,'' \emph{Proc. the Association for the Advancement of Artificial
  Intelligence (AAAI}, pp. 1004--1010, 1998.

\bibitem{Turpault_DCASE2019_01}
N.~Turpault, R.~Serizel, A.~P. Shah, and J.~Salamon, ``Sound event detection in
  domestic environments with weakly labeled data and soundscape synthesis,''
  \emph{Proc. Workshop on Detection and Classification of Acoustic Scenes and
  Events (DCASE)}, pp. 1--6, 2019.

\bibitem{Ronchini_DCASE2021_01}
F.~Ronchini, R.~Serizel, N.~Turpault, and S.~Cornell, ``The impact of
  non-target events in synthetic soundscapes for sound event detection,''
  \emph{Proc. Detection and Classification of Acoustic Scenes and Events
  (DCASE) Workshop}, pp. 115--119, 2021.

\bibitem{Mesaros_AppSci2016_01}
A.~Mesaros, T.~Heittola, and T.~Virtanen, ``Metrics for polyphonic sound event
  detection,'' \emph{Applied Sciences}, vol.~6, no.~6, p. 162, 2016.

\end{thebibliography}

\end{document}